\documentclass[12pt,a4]{article}
\usepackage{graphicx}
\begin{document}
\begin{center}
{\Large \bf Present Status of GRACE/SUSY}\\
\vspace{4 mm}

{\large \bf
FUJIMOTO Junpei$~^{a}$, ISHIKAWA Tadashi$~^{a}$, JIMBO Masato$~^{b}$,
KANEKO Toshiaki$~^{a}$, KON Tadashi$~^{c}$, KURODA Masaaki$~^{d}$
}
\vspace{4 mm}

{\large
$~^{a}~${\bf KEK,} {\it Oho, Tsukuba, Ibaraki 305-0801 Japan}\\
$~^{b}~${\bf Tokyo Management College,} {\it Ichikawa, Chiba 272-0001, Japan}\\
$~^{c}~${\bf Seikei University,} {\it Musashino, Tokyo 180-8633, Japan}\\
$~^{d}~${\bf Meiji Gakuin University,} {\it Totsuka, Yokohama 244-8539, Japan}\\
}

\end{center}

\begin{abstract}
We have developed the system for the automatic computation of cross sections,
{\tt GRACE/SUSY}~, including the one-loop calculations for processes of
the minimal supersymmetric extension of the standard model.  For an
application, we investigate the pair-production of the heavy chargino in
electron-positron collisions.
\end{abstract}

\section{Introduction}
From the theoretical point of view, it has been a promising hypothesis that
there exists a symmetry called supersymmetry (SUSY) between bosons and
fermions at the unification-energy scale~\cite{theor}.  In particular,
the minimal supersymmetric extension of the standard model (MSSM) has
been extensively studied in the past decade, because it has the simplest
structure and contains the least number of particles, and yet it is complex
enough to describe the most essential feature characteristic to any theory of
SUSY.

Since it is a broken symmetry at the electroweak-energy scale, the relic of
SUSY is expected to remain as a rich spectrum of SUSY particles, partners of
usual matter fermions, gauge bosons and Higgs scalars, named sfermions,
gauginos and higgsinos, respectively.  The quest of these new particles has
been one of the most important issues of the high-energy
physics at future colliders of sub-TeV-region or TeV-region energies.

For the simulations of the experiments, we have to calculate the cross sections
for the processes with more than three final particles because most kinds of
particles decay to two or more particles.  Several groups independently
developed computer systems which automate the perturbative calculation in the
standard model (SM) with different methods~\cite{pre-grc,comp,Den,Mad,THahn},
and also have been developing the systems of the automatic
computation in the MSSM, {\tt GRACE/SUSY}~\cite{gs,maj,gsp},
{\tt FeynArts-FormCalc}~\cite{THahns} and {\tt CompHEP}~\cite{comps}.

For more than five years, the minami-tateya group has been developing the
system of the automatic computation of the supersymmetric processes~\cite{gs,
maj}.
Compared with {\tt GRACE}~\cite{pre-grc} for the SM, {\tt GRACE/SUSY}~\cite{gsp}
for the MSSM has very complicated structure.  This is caused not only by the
complicacy of the MSSM lagrangian itself but also by the several historical
reasons in the course of the development of {\tt GRACE} and {\tt GRACE/SUSY}~.

(1) In {\tt GRACE} for the SM, the so-called Kyoto convention~\cite{kyoto} is
used
for the Feynman rule, in which, for example, the fermion propagator is given by
${{-1}\over{\gamma\cdot p-m}}$, while in {\tt GRACE/SUSY} the international
convention of the Feynman rules is adopted.  The Feynman rule of the
interaction vertex differs also in two conventions.

(2) At the first phase of the construction of {\tt GRACE/SUSY}~, we referred
to the MSSM model lagrangian presented by Hikasa~\cite{kh} and coded the
model definition files based on it.  The first paper~\cite{s23} on the MSSM
processes are computed based on these model definition files.  Later, we
noticed that the European people, in particular, the package of the generator
{\tt SUSYGEN}~\cite{sgen} defines the allowed region of the parameters $\mu$
and $\tan\beta$ differently from the Hikasa's manuscript.  This caused some
confusion in the international collaborations.  Therefore, we have decided to
have our own model lagrangian for the MSSM at hand. Kuroda has computed the
complete lagrangian of the MSSM~\cite{mk} using the European convention: namely,
the positive chargino is called a particle and the ranges of $\mu$ and
$\tan\beta$ are defined as $0\le  \tan\beta \le 1$ and
$-\infty\le \mu \le+\infty$~.

(3) In addition, the {\tt GRACE} system consists of several components,
for example, the package for the Feynman-graph
generation~\cite{Fdg}, the module for the calculation of the helicity
amptiludes {\tt CHANEL}~\cite{chan}.  We had to modify and expand them for
the construction of {\tt GRACE/SUSY}.  For the loop calculation, we also needed
to expand the package for the loop calculations in {\tt GRACE/1LOOP}~\cite{kf}.

In this paper, we provide the present status of {\tt GRACE/SUSY}~, especially
on the development of the loop calculations for the MSSM.

\section{{\tt GRACE/SUSY/1LOOP}}
For discovery experiments, we have to calculate cross sections not only for
processes of new-particle productions but also for their background processes.
For this purpose, {\tt GRACE/SUSY} is well established at the tree level, and
is widely used (for example, see \cite{higgs}).  On the other hand, we need
loop corrections for precise measurements.  The first step to apply
{\tt GRACE/SUSY} to calculations at the one-loop level was a process of the
SM-particle production \cite{gs1l}~.  Recently, we have been developing the
system for the automatic computation of the MSSM at the one-loop level
{\tt GRACE/SUSY/1LOOP}~, which is applicable to processes of the MSSM-particle
production.

We adopt the renormalization scheme of the MSSM as follows:
\begin{itemize}
\item the gauge-boson sector: the conventional approach \cite{conv}\\
(Renormalization constants of wavefunctions are introduced to unmixed bare
states and mass counterterms are introduced to mixed mass eigenstates.)
\item the Higgs sector: the Dabelstein's approach \cite{Dab}~;
 the chargino sector and the neutralino sector: the Kuroda's approach \cite{MK}
(see also \cite{Fri})\\
(Renormalization constants of wavefunctions are introduced only to unmixed
bare states.)
\item the matter-fermion sector and the sfermion sector: the Kyoto approach
\cite{kyoto}\\
(Renormalization constants of wavefunctions are introduced only to mixed
mass eigenstates.)
\end{itemize}

As an application of {\tt GRACE/SUSY/1LOOP}, we consider chargino-pair
productions in electron-positron collisions \cite{diaz,chichi}~.  For the
calculations of cross sections, we use the same input parameters as in
ref.~\cite{diaz},
 $M{\tilde{\chi}_{\rm 1}}^{+} = 150$~GeV,
 $M{\tilde{\chi}_{\rm 2}}^{+} = 420$~GeV,
 $M{\tilde{\chi}_{\rm 1}}^{\rm 0} = 75$~GeV,
 ${\rm tan}\beta = 5$,
 $A_f = M_F = 500$~GeV,
 $M{\tilde{\nu}} = 500$~GeV and
 $M_{A^0} =150$~GeV.

\begin{figure}[thb]
\centerline{\includegraphics{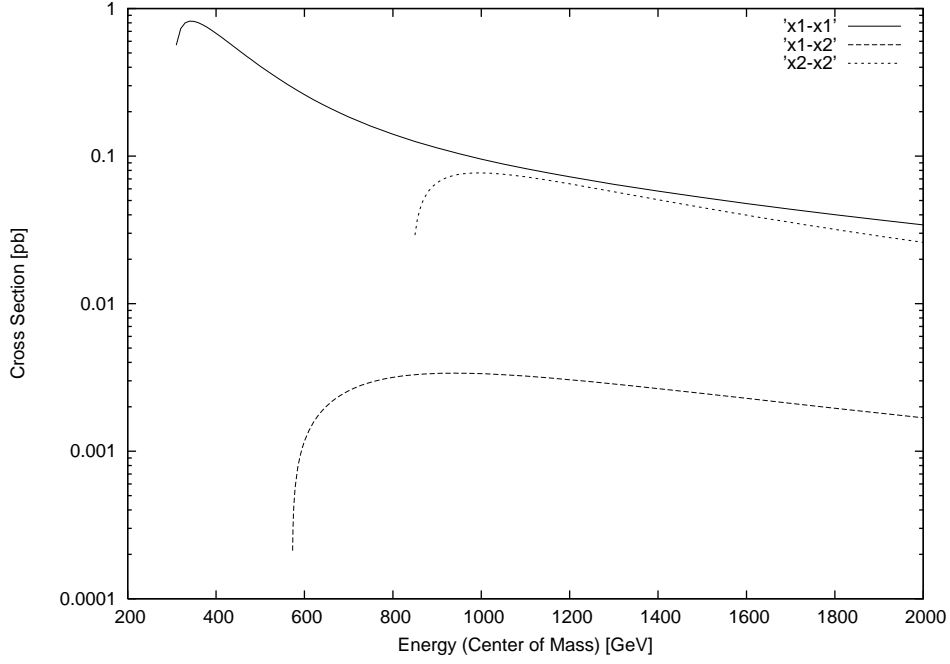}}
\caption{\label{tree} Cross-sections at the tree-level for
${e^+ e^- \rightarrow {\tilde{\chi}_{\rm i}}^{+}
{\tilde{\chi}_{\rm j}}^{-}}$~(i, j = 1 or 2).  Solid line, dashed line and
dotted line indicate cross sections for chargino1-pair production,
chargino1-chargino2 production and chargino2-pair production, respectively.}
\end{figure}

\begin{figure}[thb]
\centerline{\includegraphics{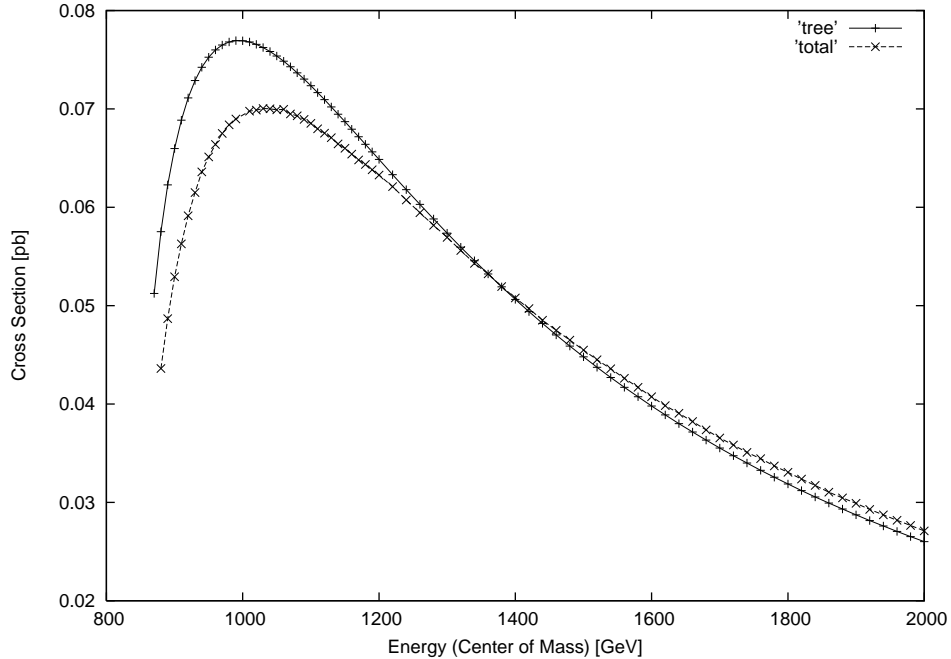}}
\caption{\label{corr} Cross-sections at the tree-level (solid line) and total
cross sections at the one-loop level (dashed line) for
${e^+ e^- \rightarrow {\tilde{\chi}_{\rm 2}}^{+}
{\tilde{\chi}_{\rm 2}}^{-}}$~.}
\end{figure}

First, we calculate cross sections at the tree-level.  The  numerical results
shown in Figure \ref{tree} indicate that the cross sections for the production
of chargino2 (heavy chargino) pair are comparable to those for the production
of chargino1 (light chargino) pair in the TeV region.  Thus we investigate the
loop calculation for the production of chargino2 pair.

\begin{table*}[htb]
\label{table:check}
\renewcommand{\tabcolsep}{2pc} 
\renewcommand{\arraystretch}{1.2} 
\center{
\begin{tabular}{@{}ll}
\hline
$C_{\rm UV}$ & 1-loop (pb) \\\hline
~~~$0$    & $-0.1913091178482273$ \\
$100$        & $-0.1913091178449565$ \\
\hline
$\lambda$    & 1-loop + soft$\gamma$ (pb) \\\hline
$1.0\times10^{-20}$    & $-7.433338646007673\times10^{-2}$ \\
$1.0\times10^{-23}$    & $-7.433338646189581\times10^{-2}$ \\
\hline
$k_{\rm c}$  & 1-loop + soft$\gamma$ + hard$\gamma$ (pb) \\\hline
$1.0\times10^{-1}$    & ${0.1374\times10^{-2}~(\pm0.000345\times10^{-2})}$\\
$1.0\times10^{-3}$    & ${0.1368\times10^{-2}~(\pm0.000473\times10^{-2})}$\\
\hline
\end{tabular}\\[2pt]
Table~1. The invariance checks of cross sections on
$C_{\rm UV}$, $\lambda$ and $k_{\rm c}$.
}
\end{table*}

For the one-loop calculations, we have to check the invariance of cross
sections varying three parameters, the UV constant ($C_{\rm UV}$), the
fictitious photon mass ($\lambda$) and the cutoff energy of the soft photon 
($k_{\rm c}$).  The invariance checks at $\sqrt{s}=1900$~GeV are shown in
Table~1.

In Figure \ref{corr}, numerical results are shown for the cross sections at
the tree-level and the cross sections at the one-loop level which include
all contributions from loop diagrams, soft-photon and hard-photon emissions.

\section{Conclusion and outlook}
We have developed the system {\tt GRACE/SUSY/1LOOP} for the automatic
computation of cross sections of the MSSM-particle production, including the
one-loop calculations For an application, we have
investigated the pair-production of the heavy chargino in electron-positron
collisions, and tuned up our system.

Remaining tasks for us are:
\begin{itemize}
\item checking {\tt GRACE/SUSY/1LOOP} with the non-linear gauge in the MSSM\\
(Checking {\tt GRACE} with the non-linear gauge has already been done in the SM
\cite{nlg}.)
\item checking {\tt GRACE/SUSY/1LOOP} for the invariance on the UV constant
with other processes
\end{itemize}

\section*{Acknowledgements}
This work was partly supported by Japan Society for Promotion of Science
under the Grant-in-Aid for Scientific Research B ( No.14340081 ).


\end{document}